\documentclass[conference]{IEEEtran}
\IEEEoverridecommandlockouts
\usepackage{blindtext}
\usepackage{cite}
\usepackage{amsmath,amssymb,amsfonts}
\usepackage{enumitem}
\usepackage[linesnumbered,ruled,vlined]{algorithm2e}
\usepackage[noend]{algpseudocode}
\usepackage{graphicx}
\usepackage{textcomp}
\usepackage{svg}
\usepackage{xcolor}
\usepackage[utf8]{inputenc}
\usepackage{lipsum}

\def\BibTeX{{\rm B\kern-.05em{\sc i\kern-.025em b}\kern-.08em
    T\kern-.1667em\lower.7ex\hbox{E}\kern-.125emX}}
\usepackage[left= 1.6 cm, right= 1.6 cm, top=1.778 cm, bottom = 2.65cm]{geometry}
\begin{document}

\title{Cooperative Caching Towards Efficient Spectrum Utilization in Cognitive-IoT Networks}


\author{\IEEEauthorblockN{Nadia Abdolkhani and Walaa Hamouda}
\IEEEauthorblockA{Department of Electrical and Computer Engineering \\
Concordia University,
Montreal,  Quebec, H3G 1M8, Canada\\
email:\{n\_abdolk, hamouda\}@ece.concordia.ca}}
\maketitle

\begin{abstract}
In cognitive Internet of Things (CIoT) networks, efficient spectrum sharing is essential to address increasing wireless demands. This paper presents a novel deep reinforcement learning (DRL)-based approach for joint cooperative caching and spectrum access coordination in CIoT networks, enabling the CIoT agents to collaborate with primary users (PUs) by caching PU content and serving their requests, fostering mutual benefits. The proposed DRL framework jointly optimizes caching policy and spectrum access under challenging conditions. Unlike traditional cognitive radio (CR) methods, where CIoT agents vacate the spectrum for PUs, or relaying techniques, which merely support spectrum sharing, caching brings data closer to the edge, reducing latency by minimizing retrieval distance. Simulations demonstrate that our approach outperforms others in lowering latency, increasing CIoT and PU cache hit rates, and enhancing network throughput. This approach redefines spectrum sharing, offering a fresh perspective on CIoT network design and illustrating the potential of DRL-guided caching to highlight the benefits of collaboration over dynamic spectrum access scenarios, elevating CIoT performance under constrained resources.
\end{abstract} 
\section{Introduction}
\IEEEPARstart{T}{he} rapid growth of smartphones, cloud computing, and the Internet of Things (IoT) has significantly increased demand for wireless spectrum, driving interest in cognitive radio (CR) for IoT networks (CIoT) to dynamically access underutilized bands \cite{Nadia_Jamming_IoTJ_2024, Nadia_Nada_SWIPT}. Cognitive radio networks (CRNs) present a promising solution to enhance spectrum efficiency and alleviate scarcity by allowing unlicensed secondary networks to opportunistically access spectrum licensed to primary users (PUs). In CRNs, CIoT agents can share spectrum through underlay, overlay, and interweave paradigms \cite{nada_survey_2023}. The overlay approach lets CIoT agents offer specific services in exchange for licensed spectrum access. However, when not offering these services, CIoT agents must vacate the spectrum upon PU access. This transition incurs delays as unlicensed users switch channels, causing potential interruptions and increased transmission latency, challenging the delay constraints of many CIoT applications.

In most studies, relaying PU data serves as a bargaining tool for accessing licensed spectrum \cite{Farooq_relaying_Access_2024, Liang_relaying_Ucom_2024, Wilson_relaying_ICCC_2024}, benefiting both primary and secondary systems. However, limited backhaul capacity can reduce these gains, especially with small base stations in IoT networks that increase spatial spectrum reuse, amplifying the need to minimize backhaul overhead. To address this, caching popular files at the network edge has been proposed to reduce backhaul load and improve user delay \cite{Abdolkhani_cache_Access_2022}. Traditional non-learning methods assume known content popularity to model request patterns, which is unrealistic \cite{Li_cache_nonlearning_2021, Yang_cache_nonlearning_2019, Yang_cache_nonlearning_2020, Nissar_cache_gametheory_2019}. Effective edge caching, however, relies on dynamically changing content popularity, which varies temporally and spatially and is often unknown in advance. 

Due to the unknown content popularity, the complex environment, and the need for dynamic decision-making, reinforcement learning (RL) emerges as a promising approach to optimize the CIoT agent’s action policy through interactive learning. Yang et al. proposed a bisection search for optimizing joint caching and power allocation to minimize backhaul delay, while Nissar et al. used a game-theoretic approach for cache-enabled cognitive device-to-device (D2D) networks, and Li et al. applied linear programming to jointly schedule caching and transmission to reduce network latency. Furthermore, \cite{Gao_Liu_coopcaching_2022} proposed a heterogeneous multi-agent deep deterministic policy gradient (MADDPG) approach in a cooperative cache-enabled CRN. Gao et al. leveraged MADDPG’s centralized training and decentralized execution to optimize content caching at the secondary base station (SBS), while retrieving uncached content from the primary base station (PBS). However, a continuous approach such as DDPG would increase computational complexity and runtime without yielding comparable performance gains in resource-constrained CIoT devices. It is important to note that multi-agent reinforcement learning approaches usually face convergence problems which impacts the stability and adaptability of the approach in dynamic environments.

Given the aforementioned gap, we propose a novel deep reinforcement learning (DRL)-based approach for jointly optimizing channel access and caching policy to maximize throughput and minimize service delay. In our overlay CIoT network, secondary devices operate by fulfilling the PU's content requests as part of our cooperative scheme, without the knowledge of their channel occupancy, transmission power, or signal/noise characteristics. In this uncertain environment, dynamic decision-making is crucial, and DRL proves highly effective, enabling CIoT devices to learn optimal strategies through continuous interaction with the environment without requiring explicit knowledge. We further introduce an innovative Upper Confidence Bound (UCB) strategy to finely balance the exploration-exploitation trade-off. Given the computational complexity of continuous action spaces and the CIoT agent's limited resources, a discrete choice is both practical and effective for CIoT networks. Thus, We formulate the DRL framework as a discrete-time model-free Markov decision process (MDP) with continuous states and discrete actions, and a novel double deep Q-network (DDQN) architecture is proposed. Our proposed DRL algorithm undergoes comprehensive convergence and performance analysis, benchmarked against alternative methods from existing literature across various test scenarios. We also evaluate the impact of small-scale fading channels, alongside a comparison between cooperative and non-cooperative scenarios, highlighting the significance of cooperation in reducing delay within CRNs. 

The rest of the paper is structured as: Section II presents the system model and problem formulation, Section III showcases the proposed DRL approach for joint caching policy and access coordination, Section IV discusses the simulation model and presents a comprehensive analysis of the paper's results, and finally, Section V concludes the paper.
\section{System Model}
Consider a time-slotted communication system over a finite horizon of \( T \) time slots, each with duration \( \tau \). In this CIoT network, a transmitter-receiver (Tx-Rx) pair shares the spectrum alongside a primary Tx-Rx pair. The primary users (PUs) occupy \( L \) slots for transmission, with the PU transmitter consistently operating at a power level \( P_p^t \) and the CIoT agent at a power level \( P_s^t \) in each time slot. When the PU transmits in the \( t \)-th slot, the PU status indicator \( \omega_p^t \) is 1; otherwise, it is 0. The primary and secondary networks operate independently, with PUs and CIoT devices interested in different content sets. Let \( \mathcal{M} = \{1, 2, \ldots, M\} \) and \( \mathcal{N} = \{1, 2, \ldots, N\} \) represent the sets of primary and secondary content, respectively. For simplicity, we assume uniform content size, although varying content sizes can be managed by dividing them into equal-sized chunks.\footnote{This assumption can be easily removed, as content with varying sizes can be divided into equal-sized chunks.}

The popularity of primary and secondary content is represented by \( Q_p \) and \( Q_s \), respectively, and is initially unknown to the agent. Content requests for both PU (denoted as \(d_{p}^t\)) and CIoT devices (denoted as \(d_{s}^t\)) follow the widely used Zipf distribution \cite{Yang_cache_nonlearning_2019}, with skew factors \( \gamma_p \) and \( \gamma_s \), and arrival rates of \( \lambda_p \) and \( \lambda_s \) requests per slot, respectively. Let \( C_s \) denote the cache capacity of the CIoT agent. Let \( B_p^t = \{B_{p1}^t, B_{p2}^t, \ldots, B_{pM}^t\} \) and \( B_s^t = \{B_{s1}^t, B_{s2}^t, \ldots, B_{sN}^t\} \) represent the cooperative caching policy of the CIoT agent for caching PU and CIoT content, respectively. Here, \( B_{pm}^t = 1 \) indicates that the \( m \)-th PU content is cached in the \( t \)-th time slot, and zero otherwise. Here, \( B_{sm}^t = 1 \) indicates that the \( m \)-th CIoT agent's content is cached in the \( t \)-th time slot, and zero otherwise. The channel power gains for the CIoT Tx-Rx pair \( g_{ss}^t \), between the PU Tx and CIoT Rx \( g_{ps}^t \), and between the CIoT Tx and PU Rx \( g_{sp}^t \) are modeled as independent and identically distributed (i.i.d.) Rayleigh fading channels. These channel power gains are assumed constant throughout each time slot.

Under the cooperative scheme, if the channel is idle and the CIoT agent's content request is cached locally, the achievable rate of the CIoT agent in the \( t \)-th time slot is given by \begin{equation}\label{eq:r_0}
 R_0^t = W \log_2\bigg(1+\frac{P_s^t g_{ss}^t}{\sigma^2}\bigg),
\end{equation} where \( \sigma^2 \) is the channel noise variance and \( W \) is the total licensed bandwidth of the PU channel. If the channel is occupied by the PU Tx in the \( t \)-th slot, and the CIoT agent has both the PU's and its own content request cached locally, the PU will allow access to a portion \( k \) of the bandwidth. Consequently, the achievable rate of the CIoT agent decreases, given by \begin{equation}
 R_1^t = \frac{W}{k} \log_2\bigg(1+\frac{P_s^t g_{ss}^t}{\sigma^2}\bigg).
\end{equation} If neither condition is met, the CIoT agent will need to offload its request from the core network.

In the studied cooperative model, the CIoT Tx seeks to maximize its rate while considering, the channel occupation, the channel access coordination with the PU, and the caching policy of both PU content and CIoT content. Therefore, the maximization of the rate of the CIoT Tx is modeled as a constrained optimization problem as
\begin{subequations}
\label{eqn:optim}
\begin{align}
   &\max_{B_{pm}^t,B_{sn}^t,I_t} \sum_{t=1}^{T} (1-I_t)\big{[}(1-\omega_{p}^t)R^t_0+\omega_{p}^tR^t_1\big{]} \label{maximization}\\
   &\text{s.t.  }   d_{p}^t \in B_{pm}^t , d_{s}^t \in B_{sn}^t\label{constraint1}\\
   &\sum_{m=1}^{M}P_{pm}^t+\sum_{n=1}^{N}P_{sn}^t\leq C_s\label{constraint2}\\
   & I_t = 0 , ~~ \forall \omega_n^t=0 ~\text{for}~ n=1,...,N\label{constraint3},
\end{align}
\end{subequations} 
where \(I_t\) is CIoT agent's decision to cache PU content (\(I_t=1\)) or not (\(I_t=0\)).

Under the non-cooperative scheme, if the channel is idle and the CIoT agent's content request is cached locally, the achievable rate of the CIoT agent in the \( t \)-th time slot is given by Eq.~\eqref{eq:r_0}. If the channel is occupied by the PU transmitter, the CIoT agent must vacate the channel and retrieve its data from the core network, as it does not have the PU content cached locally. 
\section{Optimal Joint Channel Access Coordination and caching policy strategy}
The process of learning the optimal strategy can be formulated as a Markov Decision Process (MDP) with the tuple \((\boldsymbol{S}, \boldsymbol{A}, \mathcal{P}, \boldsymbol{R}, T)\), where:
\(\boldsymbol{S}\) denotes the set of states, \(\boldsymbol{A}\) represents the set of actions, \(\mathcal{P}\) is the set of state transition probabilities, \(\boldsymbol{R}\) is the set of rewards, and \(T\) is the time step. In practical CR scenarios, determining the exact transition probabilities \(\mathcal{P}\) is challenging due to unpredictable energy and channel fading. Thus, we employ a model-free MDP, using a DRL framework to estimate rewards \(\boldsymbol{R}\) based on states \(\boldsymbol{S}\) and actions \(\boldsymbol{A}\) without requiring \(\mathcal{P}\). This simplifies the MDP tuple to \((\boldsymbol{S}, \boldsymbol{A}, \boldsymbol{R}, T)\). Below, we detail each component of the MDP tuple.

\textbf{State Space \(\boldsymbol{S}\):} The state \(s_t\) at each time slot \(t\) includes the PU occupancy indicator \(\omega_p^t\), the content request from the \((t-1)\)-th time slot \(d_p^{t-1}\) and \(d_s^{t-1}\), and the channel power gains \(g_{ps}^t\), \(g_{sp}^t\), and \(g_{ss}^t\). Therefore, the state can be expressed as \( s_t = \{ \omega_p^t, d_p^{t-1}, d_s^{t-1}, g_{ps}^t, g_{sp}^t, g_{ss}^t \}\). It is important to note that in practical scenarios, the agent does not have knowledge of content popularity or its specific content request at the \(t\)-th time slot, so these elements are excluded from the observed state space of the CIoT agent.

\textbf{Action Space \(\boldsymbol{A}\):} At each time slot \( t \), the action \( a_t \) involves deciding whether to cache the PU contents, thereby sharing the cache capacity \((I_t = 1)\), or not sharing it \((I_t = 0)\). Based on this decision \( I_t \), the agent sets the caching policies accordingly. Thus, the action is defined as \( a_t = [I_t, B_{pm}^t, B_{sm}^t] \), where \( B_{pm}^t \) and \( B_{sm}^t \) represent the caching decisions for the PU contents and the SU contents at time \( t \), respectively. For the non-cooperative scheme the \(B_{pm}^t\) will be excluded from the action space.

\textbf{Reward \(\boldsymbol{R}\):} The reward \(r_t\) depends on the achievable rate when transmitting, provided the constraints in Eq.~(\ref{eqn:optim}) are satisfied. A penalty \(-\phi\) is applied if constraints are violated. The reward function is:
\begin{equation}\label{eq:reward}
 r_t = \begin{cases} 
    R_1^t & \text{if } I_t=1, \omega_{p}^t=1, d_p^t \in B_{pm}^t, d_s^t \in B_{sn}^t, \\
    R_0^t & \omega_{p}^t=0, ,  d_s^t \in B_{sn}^t, \\
    -\phi & \text{otherwise}.
  \end{cases}
\end{equation}
For the non-cooperative scheme, the \(R_1^t\) term is excluded from the reward function.

\textbf{Time Step \(T\):} Each transition from time slot \(t\) to \(t+1\) is considered a single step, iterating through state-action pairs across \(T\) time slots.

\subsection{Double Deep Q-Network}
In the model-free MDP, the CIoT agent must estimate the state-action value without prior knowledge of \(\mathcal{P}\). Through reinforcement learning (RL), the CIoT agent can approximate the state-value function and learn a policy \(\pi\) for selecting actions based on the environment's current state. The goal is to maximize long-term cumulative reward (rate) while adhering to CR system constraints. Q-learning, an RL algorithm, focuses on estimating the expected state-action value, or Q-function:

\begin{equation}
Q^\pi (s,a) = \mathbb{E}[r_t + \beta Q^\pi(s_{t+1}, a_{t+1}) | s_t = s, a_t = a],
    \label{eqn:q_learning}
\end{equation}

where \(r_t\) is the immediate reward for action \(a_t\) in state \(s_t\), and \(\beta\) is a discount factor that balances immediate and future rewards. The SU Tx agent seeks the optimal action \(a\) that maximizes Q-value at each slot \(t\), i.e., \(a^* = \arg \max_{a \in \boldsymbol{A}} Q^\pi(s,a)\).

Deep Q-networks (DQNs) have been widely used to learn optimal actions in discrete action spaces; however, they suffer from overestimation bias, which can lead the learning agent to select overly optimistic actions, resulting in suboptimal performance. To address this, we propose a double deep Q-network (DDQN), which approximates the Q-function using a neural network to mitigate overestimation bias. 

The DDQN predicts the cumulative reward (Q-value) for each action \(a\) in a given state \(s\), updating its parameters \(\boldsymbol{\theta}\) so that \(Q^\pi(s,a; \boldsymbol{\theta}) \approx Q^\pi(s,a)\). The DDQN architecture consists of an input layer representing the state space \(\boldsymbol{S}\), two hidden layers with \(h_1\) and \(h_2\) neurons, and an output layer with \(z\) neurons. The parameters \(\boldsymbol{\theta} = \{ \textbf{W}^{(i)}, \textbf{b}^{(i)} \}\) include weights and biases across the network layers \(i = \{1, \ldots, 4\}\).

Weights are initialized using Kaiming (He) initialization, drawing from \(\mathcal{N}(0, \frac{2}{\nu_i})\), where \(\nu_i\) is the number of input neurons in layer \(i\), promoting faster convergence and improved generalization. We apply a leaky ReLU activation function \(f(x)\) to mitigate the "dying ReLU" problem, defined as:
\begin{equation}
f(x) = \begin{cases}
    x, & \text{if } x \geq 0, \\
    \alpha x, & \text{if } x < 0,
\end{cases}
\label{eqn:leaky_relu}    
\end{equation}
where \(\alpha\) adjusts the "leakiness" of the ReLU.

During training, a Target Q-network, initially identical to the Q-network, is updated less frequently to improve stability. The mean squared error (MSE) loss \(\mathcal{L}\) calculates the difference between predicted and target Q-values for a mini-batch of state-action pairs \((\textbf{s}, \textbf{a})\):
\begin{equation}
\begin{split}
    \mathcal{L}(\boldsymbol{\theta}) = \mathbb{E}\Bigg[ \bigg[ \textbf{r}_t + \gamma Q^\pi\bigg(\mathbf{s}_{t+1} &, \arg \max_{\mathbf{a}\in \boldsymbol{A}}~ Q^\pi(\mathbf{s}_{t+1}, \mathbf{a}_t; \boldsymbol{\theta}); \boldsymbol{\theta}' \bigg) \\
    &-Q^\pi(\mathbf{s}_t,\mathbf{a}_t;\boldsymbol{\theta})\bigg]^2\Bigg],
    \label{eqn:loss_fn}
\end{split}
\end{equation}
where \(\boldsymbol{\theta}'\) represents the Target Q-network's parameters. Training uses experience replay, where an experience buffer stores past interactions \((s, a, r, t)\). Once the buffer size exceeds a threshold \(\kappa\), mini-batches are sampled to decorrelate data and enhance stability.

To minimize the loss in Eq.~(\ref{eqn:loss_fn}), backpropagation calculates \(\nabla_{\boldsymbol{\theta}}\mathcal{L}(\boldsymbol{\theta}; \textbf{s}, \textbf{a})\), the gradient of the loss relative to DDQN parameters. The parameters are updated using stochastic gradient descent (SGD):
\begin{equation}
    \boldsymbol{\theta} = \boldsymbol{\theta} - \eta \nabla_{\boldsymbol{\theta}} \mathcal{L}(\boldsymbol{\theta}; \textbf{s}, \textbf{a}),
    \label{eqn:sgd}
\end{equation}
where \(\eta\) is the learning rate. We use Adam, an advanced SGD-based optimizer, for efficient updates \cite{autoencoder,icc2022}. Additionally, a dynamic learning rate scheduler adjusts \(\eta\) based on model performance, supporting efficient convergence.

\subsection{Proposed Upper Confidence Bound Strategy}
As widely recognized in the literature, there is a trade-off between exploring new actions in the action space (i.e., learning their expected rewards) and exploiting known actions that yield the highest empirical rewards. Most studies employ the $\epsilon$-greedy strategy to balance this exploration-exploitation trade-off \cite{Khalek_IoT_2024}. If the expected rewards of all actions were known, the optimal policy would simply select the action with the highest expected reward. However, for the CIoT agent to effectively explore the environment, discover optimal strategies, and balance this trade-off, we apply principles from the Upper Confidence Bound (UCB) algorithm.

To enhance exploration and identify optimal strategies, we propose a novel variant of the UCB strategy called UCB-Zipf (UCBZ). This variant adjusts Q-values based on the Zipf-like distribution of content popularity, tailored specifically to the problem context at hand. The adjusted Q-value is defined as:
\begin{equation}\label{eq:ucb}
\overline{Q}^\pi (s,a) = Q^\pi (s,a) + \frac{1}{M^{\gamma_p} \cdot N^{\gamma_s}}\sqrt{\frac{c'\ln{t}}{C_a^t}},
\end{equation}
where \(c'\) is a hyperparameter of the UCBZ algorithm, \(M\) is the number of PU content, \(N\) is the number of CIoT agent content, \(C_a^t\) represents the count of action \(a\) taken up to the \(t\)-th time slot, and \(\gamma_p\) and \(\gamma_s\) are the zipf parameters of PU and CIoT content, respectively. The UCBZ algorithm has a space complexity of \( \mathcal{O}(z) \), where \( z \) is the number of output neurons of the DDQN. Additionally, its computational complexity is \( \mathcal{O}(1) \). The space complexity of the DDQN-UCBZ is $\mathcal{O}(\mathcal{M})$ and its computational complexity per time step is $\mathcal{O}(\theta)$, where $\theta$ is the total number of parameters in the DDQN.

\begin{figure}
    \centering
    \includegraphics[width=1\columnwidth]{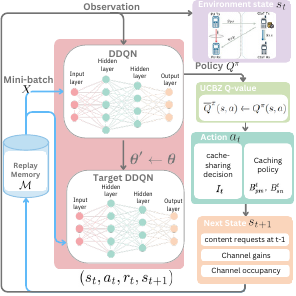}
    \caption{The proposed DDQN-UCBZ algorithm}
    \label{fig:dqn_model}
\end{figure}

This approach enables the agent to acquire essential knowledge about the environment during early training stages and gradually shift toward exploiting this knowledge for reward maximization. For the non-cooperative scheme since it will not cache the PU content then the term \(M^{\gamma_p}\) is excluded from Eq.~\ref{eq:ucb}. The training process for the DDQN-UCBZ algorithm is shown in Algorithm \ref{alg:algorithm1}. Furthermore, Fig.~\ref{fig:dqn_model} provides a detailed view of the employed DRL algorithm, including our proposed UCBZ exploration strategy.

\begin{algorithm}[t!]
\caption{The proposed DDQN-UCBZ Algorithm}\label{alg:algorithm1}
\textbf{Input:} Cognitive IoT environment simulator and parameters.

\textbf{Output:} Optimal action $a_t$ in each time slot $t$.

Initialize experience replay memory $\mathcal{M}$ with size $m$.

Initialize $B_0$, $\eta$, $\beta$, $\kappa$, and $c'$.

\For{episode= 1,...,episodes}{
  
    \For{t= 1,...,T}{
        Observe the state $s_t$;
        \If{$\mathcal{M}$ is not full}{
        
        Sample a random action $a_t$;
        }

        \Else{ 
            Adjust Q-value using (\ref{eq:ucb});
            
            Get action $a_t$ according to the policy of adjusted Q-value;           
            }

            Get the reward $r_t$ using (\ref{eq:reward}); 
            
            observe the next state $s_{t+1}$ ;
        
            Store $\mathcal{M}\leftarrow(s_t,a_t,r_t,s_{t+1})$;
        
            Update action count, $C_a^t \leftarrow C_a^t + 1$; 

            Sample a mini-batch $X$ from $\mathcal{M}$ ;

            Predict Target Q-values using 
            
            $\textbf{r}_t + \beta \max_{\textbf{a}\in \boldsymbol{\mathcal{A}}}~ Q^\pi(\textbf{s}_{t+1},\textbf{a}_{t+1};\boldsymbol{\theta}')$ ;

            Predict Q-values using $Q^\pi(\boldsymbol{s},\boldsymbol{a};\boldsymbol{\theta})$;

            Calculate the loss in (\ref{eqn:loss_fn});
            
            Update $\boldsymbol{\theta}$ of DDQN online;
            
            \If{episode*t \text{mod} $\kappa$ = 0}
            {
            Update $\boldsymbol{\theta}'$ of Target DDQN online as $\boldsymbol{\theta}' \leftarrow \boldsymbol{\theta}$ ;
             }

}
         Update $\epsilon$ and the state $s_{t+1} = s_t$;
         
         Update $\eta$ using scheduler ;
}
\end{algorithm}
\section{Simulation Model and Results}
In this section, we analyze the performance of our proposed DRL approach in a CIoT network with a Tx-Rx pair sharing a channel with a PU Tx-Rx pair in a non-line-of-sight (NLoS) environment. We consider a channel noise variance of $\sigma^2=10^{-3}$ and a path-loss exponent of \( \alpha = 4 \). The channel power gains $g_{ss}^t$ and $g_{sp}^t$ follow an exponential distribution with a mean of 0.1 and 0.2 respectively. Without loss of generality, we consider $g_{sp}^t= g_{ps}^t$. Time-slotted transmissions are considered over \( T = 30 \) slots (each lasting \( \tau = 1 \, \text{s} \)), with the PU Tx using \( L = 26 \) slots at \( P_p = 0.2 \, \text{W} \) and \( P_s = 0.1 \, \text{W} \). The DDQN architecture has four layers with \( j = 6 \), \( h_1 = 512 \), \( h_2 = 128 \), and \( z = 105 \) neurons, updating the target network every 200 iterations. The training uses an initial learning rate of \( 4 \times 10^{-4} \), halved every 500 episodes, over 2500 episodes with mini-batches of 100 frames. Constraint violations incur a penalty \( \phi = 7 \). The Target DDQN is updated every 200 iterations. The replay buffer holds \( \kappa = 333 \) experiences, with a discount factor \( \beta = 0.99 \) and UCBZ hyperparameter \( c' = 2.5 \). Furthermore, we consider \(M=5\), \(N=5\), \(C_s=4\), \(\gamma_p=0.8\), \(\gamma_s=0.6\), \(k=2\), \(\lambda_p=1\), and \(\lambda_s=1\).

In our evaluation, we employ performance metrics such as average sum rate (ASR), average delay, and average CIoT cache-hit rate. These values are the weighted moving average of the total achievable metrics. The cache-hit rate is calculated as the number of times the CIoT agent's request is serviced from the local cache divided by the total number of CIoT agent's content request.

To assess the effectiveness of our proposed DDQN-UCBZ approach, we compare the system performance with the following:
\begin{enumerate}[label=(\roman*)]
\item the non-cooperative caching scheme that does not share its cache capacity with the licenced user and has to vacate the spectrum when the PU is transmitting.
    \item The $\epsilon$-greedy strategy, which is widely adopted in the literature to balance exploration and exploitation. 
    Using the $\epsilon$-greedy strategy, the CIoT agent selects an action aiming to maximize the estimated Q-value (exploitation) with a probability of $1-\epsilon$, while opting for a random action (exploration) with a probability of $\epsilon$. 
    \item the least recently used (LRU) caching algorithm that always keeps the most recently requested content item in the cache. LRU is a common cache replacement strategy used in caching literature.
\end{enumerate}

\begin{figure}[t!]
    \centering
\includegraphics[width=1\columnwidth]{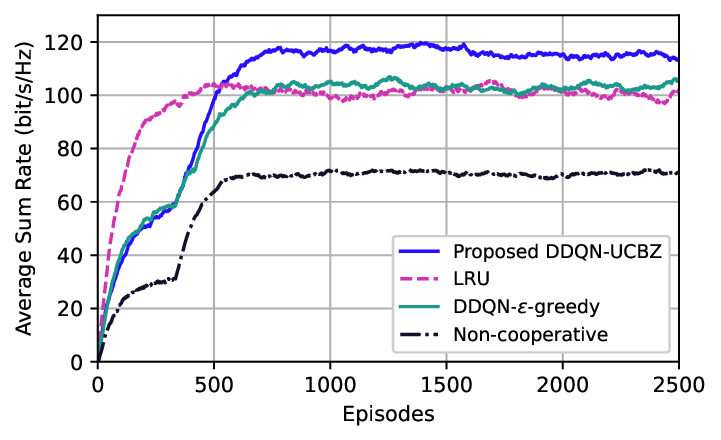}
    \caption{The CIoT agent's ASR during episodes of training.}
    \label{fig:benchmarking}
\end{figure}

In Fig.~\ref{fig:benchmarking}, we illustrate the CIoT agent's ASR across training episodes for various strategies. At the start of training, the agent accumulates experience in its replay buffer, with learning commencing after episode 333. Our proposed DDQN-UCBZ strategy clearly outperforms all other strategies following convergence. By comparing with the non-cooperative strategy, we demonstrate that the cooperative scenario significantly improves throughput, resulting in reduced service delay and highlighting the impact of cooperation. This improvement is due to the CIoT agent having more transmission opportunities by caching and providing PU content, thereby gaining access to the licensed spectrum. The non-cooperative strategy achieves the lowest ASR, as the agent, lacking access to the PU's spectrum, is forced to vacate the channel when the PU is active, limiting transmission opportunities. Meanwhile, the LRU algorithm and the DDQN-\(\epsilon\)-greedy strategies yield similar ASR levels, suggesting that the performance boost in our approach is largely due to our novel UCBZ variant. This finding indicates that not all learning methods achieve optimal performance, and given resource constraints, LRU may be a more efficient choice than DDQN-\(\epsilon\)-greedy.

\begin{figure}[t!]
    \centering
\includegraphics[width=1\columnwidth]{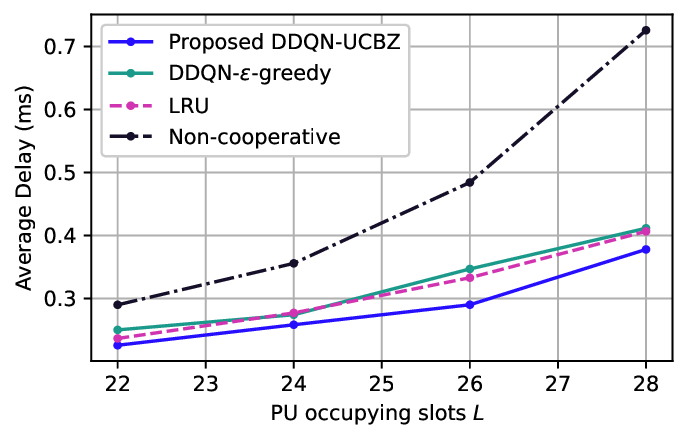}
    \caption{The CIoT agent's service delay across different values of $L$ the number of PU occupying time slots.}
    \label{fig:delay_pu_slots}
\end{figure}

\begin{figure}[t!]
    \centering
\includegraphics[width=1\columnwidth]{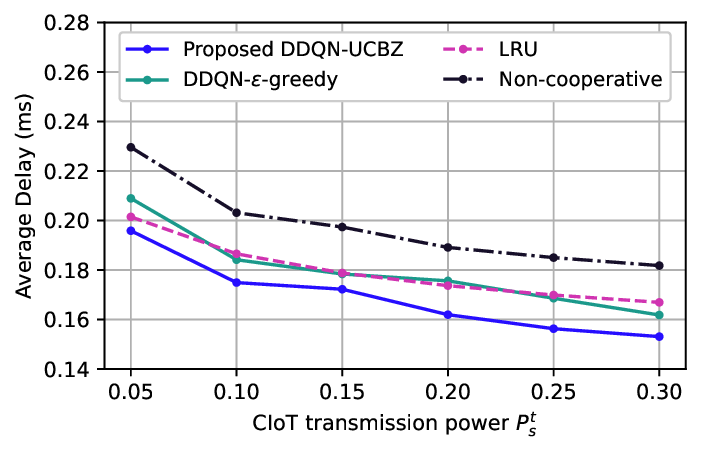}
    \caption{The CIoT agent's service delay across different CIoT transmission power $P_s^t$.}
    \label{fig:delay_p_su}
\end{figure}

\begin{figure}[t!]
    \centering
\includegraphics[width=1\columnwidth]{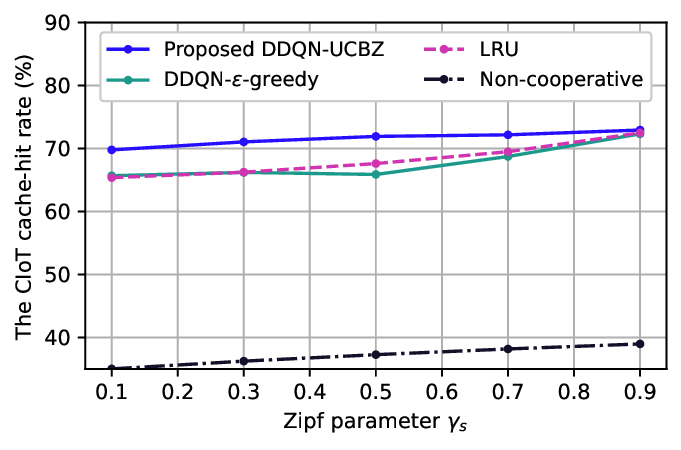}
    \caption{The CIoT agent's cache-hit rate across different zipf parameters $\gamma_s$.}
    \label{fig:su_hit_rate}
\end{figure}

In Fig.~\ref{fig:delay_pu_slots}, we present the CIoT agent's service delay across different numbers of PU-occupied slots \(L\) for various strategies. As shown, our proposed DDQN-UCBZ strategy achieves the lowest delay compared to other strategies, while the non-cooperative strategy results in the highest delay. Both the LRU and DDQN-\(\epsilon\)-greedy strategies yield similar delay levels, performing better than the non-cooperative approach but falling short of our strategy. The figure also shows that as \(L\) increases, service delay grows. This is because, with more slots occupied by the PU, the CIoT agent has fewer opportunities for full-spectrum transmission. Even when accessing the PU's spectrum by providing cached data, the agent is limited to a portion of the spectrum, which in turn increases transmission delay.
In Fig.~\ref{fig:delay_p_su}, we demonstrate the effect of varying the CIoT agent's transmit power \( P_s^t \) on the service delay for various strategies. As shown in the figure, increasing the transmit power leads to a decrease in service delay due to higher transmission rates. Furthermore, our proposed DDQN-UCBZ strategy outperforms the other strategies by achieving the lowest delay across all power levels, while the non-cooperative strategy performs the worst, exhibiting the highest delay. This poor performance is attributed to the limited transmission opportunities caused by the PU's restrictions.
We also observe that the LRU strategy and the DDQN-\(\epsilon\)-greedy strategy exhibit similar performance, which further confirms the effectiveness of our novel UCBZ variant and its superior ability to balance the exploration-exploitation trade-off.
In Fig.~\ref{fig:su_hit_rate}, we show the effect of varying the Zipf parameter \(\gamma_s\) on the cache-hit rate of the CIoT agent for different strategies. As seen in the figure, increasing \(\gamma_s\) leads to a slight rise in cache-hit rates across all strategies. This is because a higher \(\gamma_s\) increases the likelihood of users requesting the most popular content more frequently, thus improving the probability of a cache hit by storing the most popular items. This explains how at higher values of \(\gamma_s\), the proposed DDQN-UCBZ, LRU, and DDQN-\(\epsilon\)-greedy strategies achieve similar cache-hit rates. Furthermore, our proposed DDQN-UCBZ strategy consistently outperforms the other strategies across various \(\gamma_s\) values. Notably, even in the worst-case scenario where \(\gamma_s = 0.1\) (when content popularity and request probabilities are relatively uniform), the DDQN-UCBZ strategy still achieves the highest cache-hit rate compared to the alternatives. The non-cooperative strategy, in contrast, yields the lowest hit rate. This can be attributed to the limited transmission opportunities for the CIoT agent in this strategy, as it must vacate the spectrum when the PU is active, thus reducing the chances to deliver cached content.

\section{Conclusions}
In this paper, we introduced a novel DRL algorithm tailored to help CIoT networks effectively balance the exploration-exploitation trade-off, optimizing action selection to maximize ASR and minimize network delay. This algorithm empowers the CIoT agent to coordinate spectrum access and facilitate cooperation by caching licensed user requests. Our results demonstrate that the proposed DDQN-UCBZ strategy achieves these objectives with significant success, outperforming existing benchmarks and underscoring the importance of adapting DRL techniques to specific system dynamics. We validated the algorithm's convergence, highlighting its potential to enhance CIoT network performance even in challenging environments. These findings emphasize the role of learning algorithms in dynamic settings and suggest promising directions for future research.
Future work will aim to expand the CIoT system model, exploring methods to adjust the model and learning parameters as the state and action spaces grow. This will likely require strategies for dimensionality reduction to maintain computational efficiency.are

\bibliography{ref.bib}

@ARTICLE{nada_survey_2023,
  author={Abdel Khalek, Nada and Tashman, Deemah H. and Hamouda, Walaa},
  journal={IEEE Commun. Surv. Tutor.}, 
  title={{Advances in Machine Learning-Driven Cognitive Radio for Wireless Networks: A Survey}}, 
  year={2023},
  volume={},
  number={},
  pages={1-1},
  doi={10.1109/COMST.2023.3345796}}

@ARTICLE{Nadia_Jamming_IoTJ_2024,
  author={Abdolkhani, Nadia and Khalek, Nada Abdel and Hamouda, Walaa},
  journal={IEEE Internet Things J.}, 
  title={Deep Reinforcement Learning for EH-Enabled Cognitive-IoT Under Jamming Attacks}, 
  year={2024},
  volume={},
  number={},
  pages={1-1},
  doi={10.1109/JIOT.2024.3457012}}

@ARTICLE{Abdolkhani_cache_Access_2022,
  author={Abdolkhani, Nadia and Eslami, Mohsen and Haghighat, Javad and Hamouda, Walaa},
  journal={IEEE Access}, 
  title={Optimal Caching Policy for D2D Assisted Cellular Networks With Different Cache Size Devices}, 
  year={2022},
  volume={10},
  number={},
  pages={99353-99360},
  doi={10.1109/ACCESS.2022.3206813}}

@ARTICLE{Farooq_relaying_Access_2024,
  author={Farooq, Muhammad and Khan, Anwar and Alhussein, Musaed and Mahmood, Hasan and Aurangzeb, Khursheed and Bhatia Khan, Surbhi},
  journal={IEEE Access}, 
  title={Outage Analysis of a Cognitive Radio System With Nakagami-m Fading and Cooperative Decode and Forward Relaying}, 
  year={2024},
  volume={12},
  number={},
  pages={144565-144578},
  doi={10.1109/ACCESS.2024.3472895}}

@INPROCEEDINGS{Wilson_relaying_ICCC_2024,
  author={Wilson, Shadrack Moses and Li, Qiang and You, Zishuo and Ge, Xiaohu},
  booktitle={Proc. IEEE Int. Conf. Commun. China (ICCC)}, 
  title={On Cooperative Spectrum Sharing with Bidirectional Relaying Networks}, 
  year={2024},
  volume={},
  number={},
  pages={1763-1768},
  doi={10.1109/ICCC62479.2024.10681731}}

@INPROCEEDINGS{Liang_relaying_Ucom_2024,
  author={Liang, Linlin and Tian, Zongkai and Huang, Haiyan and Zhang, Nina and Zhang, Dehua and Li, Yue},
  booktitle={Proc. Int. Conf. Ubiquitous Commun. (Ucom)}, 
  title={Performance Analysis of Overlay Cognitive Cooperative Communication Based on RSMA}, 
  year={2024},
  volume={},
  number={},
  pages={6-10},
  doi={10.1109/Ucom62433.2024.10695841}}

@ARTICLE{Yang_cache_nonlearning_2020,
  author={Yang, Jiachen and Ma, Chaofan and Man, Jiabao and Xu, Huifang and Zheng, Gan and Song, Houbing},
  journal={Tsinghua Science and Technology}, 
  title={Cache-enabled in cooperative cognitive radio networks for transmission performance}, 
  year={2020},
  volume={25},
  number={1},
  pages={1-11},
  doi={10.26599/TST.2018.9010137}}

@ARTICLE{Li_cache_nonlearning_2021,
  author={Li, Changkun and Chen, Wei and Letaief, Khaled B.},
  journal={IEEE Trans. Commun.}, 
  title={Joint Scheduling of Proactive Caching and On-Demand Transmission Traffics Over Shared Spectrum}, 
  year={2021},
  volume={69},
  number={12},
  pages={8319-8334},
  doi={10.1109/TCOMM.2021.3111631}}

@ARTICLE{Yang_cache_nonlearning_2019,
  author={Yang, Jiachen and Xu, Huifang and Zhang, Juping},
  journal={IEEE Commun. Lett.}, 
  title={Exploiting Secondary Caching for Cooperative Cognitive Radio Networks}, 
  year={2019},
  volume={23},
  number={1},
  pages={124-127},
  doi={10.1109/LCOMM.2018.2877383}}

@INPROCEEDINGS{Nissar_cache_gametheory_2019,
  author={Nissar, Basma and El Ouadrhiri, Ahmed and El Kamili, Mohamed},
  booktitle={Proc. Int. Wireless Commun. .Mobile Comput. Conf. (IWCMC)}, 
  title={An Evolutionary Game-Theoretic Approach for Cache-Enabled Cognitive D2D Networks}, 
  year={2019},
  volume={},
  number={},
  pages={442-448},
  doi={10.1109/IWCMC.2019.8766570}}

@ARTICLE{Gao_Liu_coopcaching_2022,
  author={Gao, Ang and Liu, Hengtong and Hu, Yansu and Liang, Wei and Ng, Soon Xin},
  journal={IEEE Commun. Lett.}, 
  title={Cooperative Cache in Cognitive Radio Networks: A Heterogeneous Multi-Agent Learning Approach}, 
  year={2022},
  volume={26},
  number={5},
  pages={1032-1036},
  doi={10.1109/LCOMM.2022.3151877}}

@INPROCEEDINGS{icc2022,  author={Abdel Khalek, Nada and Hamouda, Walaa},  booktitle={Proc. IEEE Int. Conf. Commun.},   title={{Intelligent Spectrum Sensing: An Unsupervised Learning Approach Based on Dimensionality Reduction}},   year={2022},  volume={},  number={},  pages={171-176},  doi={10.1109/ICC45855.2022.9839170}}

@ARTICLE{autoencoder,  author={J. {Xie} and J. {Fang} and C. {Liu} and L. {Yang}},  journal={IEEE Trans. Veh. Tech.},   title={{Unsupervised Deep Spectrum Sensing: A Variational Auto-Encoder Based Approach}},   year={2020},  volume={69},  number={5},  pages={5307-5319},}

@ARTICLE{Khalek_IoT_2024,
  author={Khalek, Nada Abdel and Abdolkhani, Nadia and Hamouda, Walaa},
  journal={IEEE Internet Things J.}, 
  title={Deep Reinforcement Learning for Joint Power Control and Access Coordination in Energy Harvesting CIoT}, 
  year={2024},
  volume={11},
  number={19},
  pages={30833-30846},
  doi={10.1109/JIOT.2024.3416371}}

@ARTICLE{Nadia_Nada_SWIPT,
  author={Abdolkhani, Nadia and Khalek, Nada Abdel and Hamouda, Walaa and Dayoub, Iyad},
  journal={IEEE Communications Letters}, 
  title={Deep Reinforcement Learning for Joint Time and Power Management in SWIPT-EH CIoT}, 
  year={2025},
  volume={},
  number={},
  pages={1-1},
  doi={10.1109/LCOMM.2025.3536182}}
\bibliographystyle{IEEEtran}
\end{document}